%% file: output.tex
\documentclass{article} 
\usepackage{iclr2025_conference,times}

\input{math_commands.tex}

\usepackage{hyperref}
\usepackage{url}
\usepackage{booktabs}
\usepackage{gensymb}
\usepackage{makecell}
\usepackage{graphicx}

\title{BOston Neonatal Brain Injury Data for Hypoxic
Ischemic Encephalopathy (BONBID-HIE): II. 2-year Neurocognitive Outcome and NICU Outcome}

\usepackage{authblk}

\author[1,2,\#]{Rina Bao}
\author[1,2,\#]{Yangming Ou}
\affil[1]{Boston Children's Hospital, Boston, MA, USA}
\affil[2]{Harvard Medical School, Boston, MA, USA}

\iclrfinalcopy 
\begin{document}

\flushbottom
\maketitle

\begin{abstract}
Hypoxic Ischemic Encephalopathy (HIE) affects approximately 1-5/1000 newborns globally and leads to adverse neurocognitive outcomes in 30\% to 50\% of cases by two years of age. Despite therapeutic advances with Therapeutic Hypothermia (TH), prognosis remains challenging, highlighting the need for improved biomarkers. This paper introduces the second release of the Boston Neonatal Brain Injury Dataset for Hypoxic-Ischemic Encephalopathy (BONBID-HIE), an open-source, comprehensive MRI and clinical dataset featuring 237 patients, including NICU outcomes and 2-year neurocognitive outcomes from Massachusetts General Hospital and Boston Children’s Hospital.

\end{abstract}


\section*{Background \& Summary}

HIE is a clinical syndrome due to a lack of blood flow and oxygen to the brain. It affects around 1-5/1000 newborns globally~\citep{graham2008systematic,lee2013intrapartum}. Despite advancements in Therapeutic Hypothermia (TH), the prognosis for many infants remains
challenging, with 35\%–50\% suffering adverse neurocognitive outcomes by 2 years of age~\citep{shankaran2005whole,azzopardi2009moderate,gluckman2005selective}. Therefore, 63 of the 130 ongoing HIE-related trials worldwide are testing whether new therapies~\citep{laptook2017effect,shankaran2017effect,liu2006clinical,potter2018behavioral,nunez2019topiramate,liang2019comparative,cotten2014feasibility} can supplement TH
and further reduce adverse outcomes. However, therapeutic innovation is slow and inconclusive, for 1) before therapy, patients at high risk of developing adverse outcomes cannot be identified; 2) after therapy, outcomes cannot be measured until age 2 years~\citep{laptook2021limitations}. Both issues point to a lack of a neonatal biomarker that can predict adverse 2-year outcomes. 
To facilitate the development of biomarkers in HIE study, we present BOston Neonatal Brain Injury Dataset for Hypoxic Ischemic Encephalopathy (BONBID-HIE), an open-source, comprehensive, and representative MRI and clinical dataset for HIE. This paper introduces the second part of the BONBID-HIE data. This release contains raw and derived diffusion parameter maps, as well as NICU outcome and 2-year outcome with 237 patients. Our data was from Massachusetts General Hospital (MGH) and Boston Children's Hospital (BCH). It includes MRIs from different scanners (Siemens 3T and GE 1.5T), different MRI protocols, and from patients of different races/ethnicities and ages (0-14 days postnatal age). Part I of our data release~\citep{bao2023boston} focuses on lesion detection, while Part II (this paper) is focus on clinical, treatment, and neurologic outcome data for further developing prognostic biomarkers.

\section*{Methods}

\subsection*{{Overview}}



\subsection*{Retrospective Data Collection}
Data was retrospectively collected from MGH and BCH. Inclusion criteria were: (1) term-born (at physician discretion) (2) clinical diagnosis of HIE; (3) no comorbidities such as hydrocephalus or congenital syndromes; and (4) high-quality MRI acquired in Day 0-14 after birth. Exclusion criteria were: (1) excessive motion artifacts or missing images; or (2) primary perinatal stroke, focal artery ischemic stroke, or hemorrhage. 

Clinical and demographic data were extracted from electronic health records (EHRs). MRIs were conducted using either a GE 1.5T Signa scanner or Siemens 3T TrioTim or PrismaFit scanners. Apparent diffusion coefficient (ADC) maps were generated by the scanners themselves, using Syngo software for Siemens scanners~\citep{forman2016compressed} and the Advantage Windows Workstation for GE scanners~\citep{provenzale2007diffusion,dudink2007fractional}. For further MRI details, please refer to our Part I paper~\citep{bao2023boston}.

\section*{Data Records}
{The dataset is available on Zenodo (\url{https://zenodo.org/records/13690270}). All data has been made publicly available under the CC-BY-NC-ND license (\url{https://creativecommons.org/licenses/by-nc-nd/4.0/deed.en}). }
\subsection*{Dataset Characteristics}

\input{tables/cohort}

\subsection*{Clinical Information}
Table~\ref{tab:cohort}A lists the demographics and clinical characteristics of mothers and neonates. Maternal information includes demographics (age at delivery, race), birth mode (C-section or vaginal), and complications during pregnancy and delivery. Neonatal information includes demographics (age at MRI scan, gestational age at birth, birth weight, head circumference, sex), birth conditions (1/5/10-minute APGAR scores, lowest pH value in umbilical cord), treatment (hypothermia or not), and complications in the neonatal intensive care unit (NICU), including seizure (yes/no), length of stay (in days), the use of endotracheal tube (ETT, yes/no), and the administration of total parenteral nutrition (TPN, yes/no). In each row, we also listed the number of patients who had such information available.

\subsection*{MRI Information}

\begin{figure}[htb]
    \centering
\includegraphics[width=0.7\textwidth]{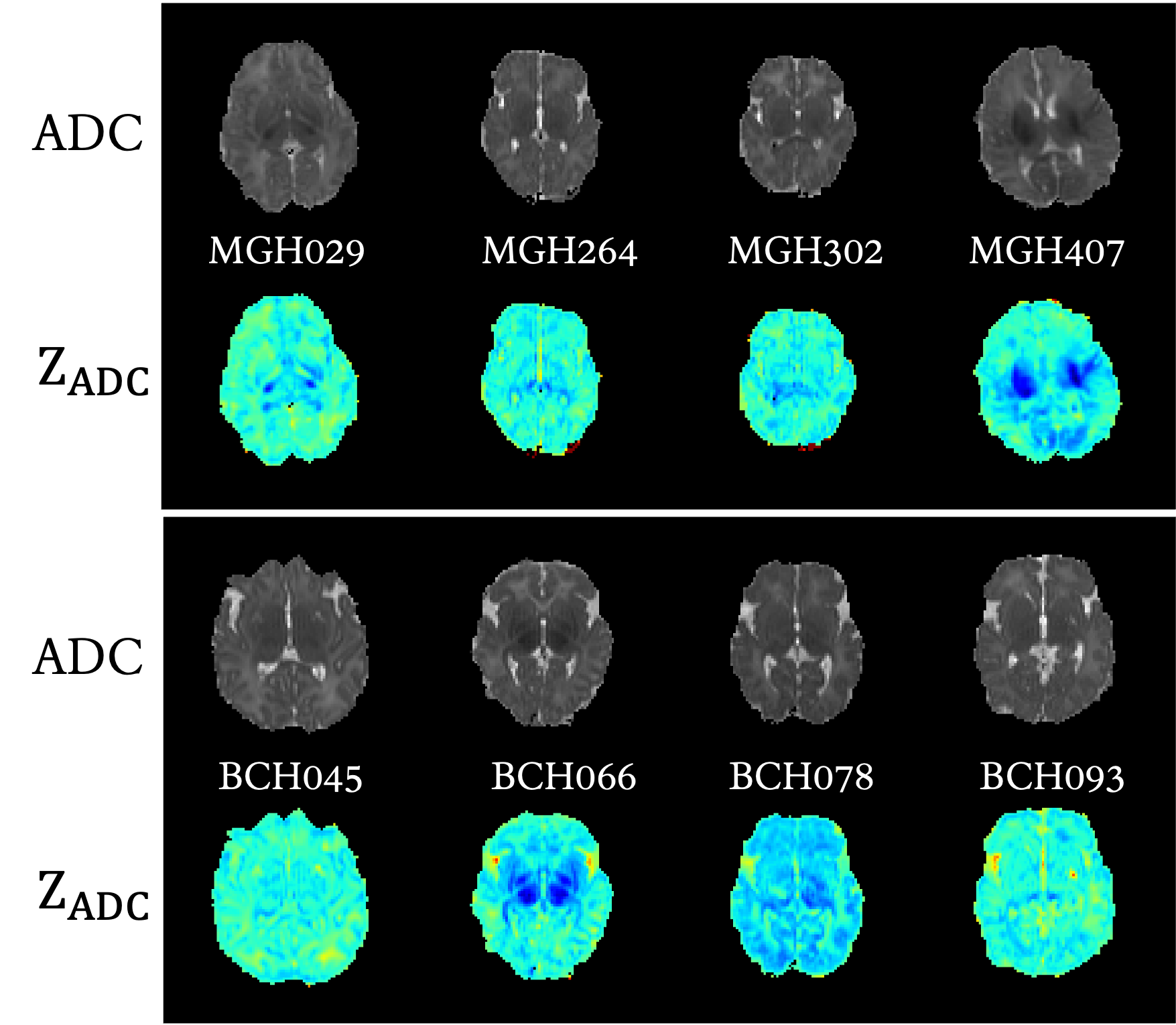}
    \caption{Representative cases for 4 HIE patients from MGH and BCH. For each patient, in the upper panel: apparent diffusion coefficient (ADC) maps; in the bottom panel: $Z_{ADC}$ maps.}
    \label{fig:visshow}
\end{figure}

We provide ADC and $Z_{ADC}$ for HIE outcome predictions, such information has been verified crucial for HIE outcome predictions. Figure~\ref{fig:visshow} shows the ADC map, $Z_{ADC}$ map from MGH and BCH. For details of ADC and $Z_{ADC}$, please refer our Part.I paper~\citep{bao2023boston}.

\input{tables/MRI}
Due to not all outcomes being collected with MRIs, we list the distribution of MRIs with NICU outcomes and 2-year outcomes in Table~\ref{tab:MRI}.

\subsection*{NICU Outcome}
Table~\ref{tab:cohort}B lists the NICU outcome which represents the patient status at NICU discharge. 

\subsection*{2-year Neurocognitive Outcome}
Table~\ref{tab:cohort}C lists characteristics of 2-year neurocognitive outcomes. Outcome is defined as normal versus adverse, according to clinical criteria and NRN recommendations~\citep{laptook2017effect,shankaran2017effect}. An adverse outcome is defined if the Bayley (version III) cognitive score$<85$ in any of the Bayley-III domains, GMFCS level between 2 to 5, or blindness/hearing impairment. Otherwise, the patient had normal outcomes. Table~\ref{tab:cohort}C lists characteristics of Bayle version III scores (BSID-III scores).

\subsection*{Data structure and file formats} 

\begin{figure}[!h]
\centering
\includegraphics[width=0.9\textwidth]{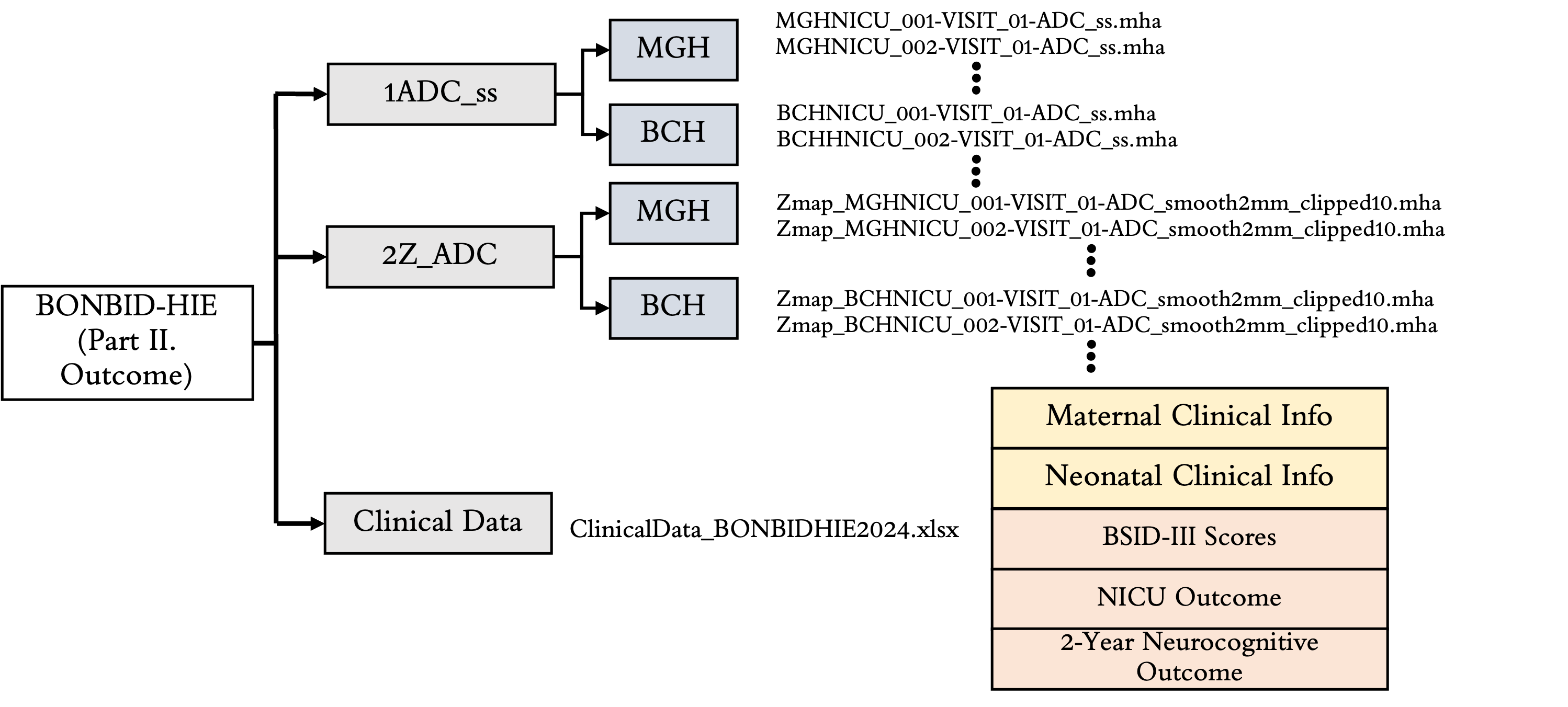}
    \caption{Folder structure of the BONBID-HIE dataset (Part II. NICU outcome and neurocognitive outcome).}
    \label{fig:dataformat}
\end{figure}

The data is organized in the format shown in Figure~\ref{fig:dataformat}. BONBID-HIE provides, per patient: (i) 1ADC\_ss: skull stripped ADC map; (ii) 2Z\_{ADC}: $Z_{ADC}$ map; (iii) 3LABEL: expert lesion annotations; and (iv) clinical data: clinical variables as written in Table~\ref{tab:cohort}A.  (v) NICU outcome: status when discharged at NICU. (vi): 2-year outcome: 2-year neurocognitive outcome, noted that patients deceased before 2-year appointment are marked as adverse outcome.

\section*{Acknowledgements} 
This work was supported, in part, by NIH R03HD104891, R21NS121735, and R61NS126792.

\section*{Competing interests} 
None.

\newpage
\bibliography{ref}
\bibliographystyle{iclr2025_conference}

\end{document}

%% file: math_commands.tex
\usepackage{amsmath,amsfonts,bm}

\def\eqref#1{equation~\ref{#1}}

\def\1{\bm{1}}

\DeclareMathAlphabet{\mathsfit}{\encodingdefault}{\sfdefault}{m}{sl}
\SetMathAlphabet{\mathsfit}{bold}{\encodingdefault}{\sfdefault}{bx}{n}



%% file: tables/cohort.tex
\begin{table}[!b]
\centering
\caption{Cohort characteristics (N=237) in MGH and BCH}\label{tab:cohort}
\resizebox{0.9\linewidth}{!}
{
\begin{tabular}{l l l }
\toprule
\textbf{A. Demographics and Clinical Characteristics}\\
\midrule
\bf{Maternal Information}\\
\midrule
Maternal age at delivery (years) &$30.30 \pm 6.04$&N=236\\
Race& \shortstack{White (117), Black or African American (13), Hispanic or Latino (18), \\ Multi Race (5), Unknown (69), Other (15)}& N=237 \\
Delivery& C-section (143), Vaginal (94)& N=237\\
Antepartum hemorrhage & Yes (43), No (194) & N=237 \\
Thyroid dysfunction & Yes (12), No (225) & N=237\\
Pre-eclampsia & Yes (10), No (227) & N=237\\
Fetal decels& Yes (153), No (84) & N=237\\
Shoulder dystocia & Yes (11), No (226) & N=237\\
Chorioamnionitis & Yes (26), No (206) & N=232\\
Emergency c-section & Yes (127), No (104) & N=231\\
\midrule
\bf{Neonatal Information}\\
\midrule
Age at scan (days) & $3.47\pm2.40$& N=232 \\
Gestational age at birth (weeks)&$39.08\pm1.94$&N=237\\
Birth weight (g)&$3294.06\pm603.91$&N=144\\
Infant head circumference (cm)&$34.28 \pm 1.70$&N=165\\
Sex & Male (137), Female (100) & N=237\\
1-minute APGAR scores &$1.89 \pm 1.80$&N=237\\
5-minute APGAR scores &$3.98 \pm 2.35$&N=235\\
10-minute APGAR scores&$5.03 \pm 2.36$&N=207\\
Lowest pH value in umbilical cord  & $6.99\pm0.22$& N=225\\
Therapeutic hypothermia before MRI? & Yes (171), No (66)& N=237\\
Endotracheal tube (ETT) in NICU & Yes (168), No (59) & N=227\\
Total parenteral nutrition (TPN) in NICU & Yes (209), No (27) & N=236\\
Seizures NICU & Yes (115), No (122) & N=237\\
Length of stay in NICU (days) & $11.68\pm9.86$& N=237\\
\midrule
\textbf{B. NICU Outcome}&Deceased (33),
Survived (204)&N=237\\

\midrule
\textbf{C. 2-year Neurocognitive Outcome}&
Adverse (103), Normal (104)& N=207\\
\midrule
Developmental Delay & Yes (61), No (173)&N=173\\
Cerebraipalsy&Yes (26), No (147)&N=173\\
Visual impairment&Yes (12), No (159)&N=171\\
Hear impairment&Yes (14), No (157)&N=171\\
\midrule
\textbf{D. BSID-III Scores}
\\
\midrule
age at test (months)&$21.41\pm2.79$&N=64\\
Raw score - cognition&$56.90\pm7.28$&N=21\\
Composite score - cognition&$95.51\pm17.47$&N=61\\
Percentile - cognition&$45.00\pm26.09$&N=60\\
Scaled score - cognition&$8.68\pm2.15$&N=19\\
Raw score - language total&n/a\\
Raw score - language (receptive communication)&$21.71\pm4.24$&N=21\\ 
Raw score - language (expressive communication)&$23.71\pm5.24$&N=21\\
Composite score - language&$93.20\pm13.14$&N=46\\
Percentile - language&$38.06\pm26.94$&N=47\\
Raw score - motor total&n/a\\
Raw score - motor (fine motor)&$35.76\pm3.18$&N=21\\
Raw score - motor (gross motor)&$50.20\pm4.52$&N=20\\
Scaled score - motor (fine motor)&$10.04\pm2.96$&N=27\\
Scaled score - motor (gross motor)&$8.53\pm1.73$&N=19\\
Composite score (motor)&$94.14\pm15.04$&N=50\\
Percentile - motor&$41.44\pm22.56$&N=50\\
\bottomrule
\end{tabular}
}
\end{table}

%% file: tables/MRI.tex
\begin{table}
\centering
\caption{Statistics for Site Data and HIE Outcome with MRIs}

\resizebox{0.65\textwidth}{!}{
\begin{tabular}{c|c|c|c|c|c|c|c}
\toprule
\multicolumn{4}{c|}{NICU Outcome with MRIs} & \multicolumn{4}{c}{2-year Outcome with MRIs} \\
\cmidrule(lr){1-4} \cmidrule(lr){5-8}
Site & Total & Survived & Deceased & Site & Total & Normal & Adverse \\
\midrule
BCH & 72 & 65 & 7 & BCH & 72 & 30 & 42 \\
MGH & 107 & 94 & 13 & MGH & 84 & 53 & 31 \\
\bottomrule
\end{tabular}
}\label{tab:MRI}
\end{table}